\documentclass[10pt,aps,twocolumn,prl,notitlepage,amssymb,amsmath,floatfix,nofootinbib,superscriptaddress]{revtex4-1}

\usepackage{epsfig}
\usepackage{bm}
\usepackage{amssymb}
\usepackage{amsmath}
\usepackage{color}
\usepackage{xcolor}
\usepackage[colorlinks,linkcolor=blue,anchorcolor=black,citecolor=blue]{hyperref}
\usepackage{multirow}

\allowdisplaybreaks[4]

\mathchardef\mhyphen="2D

\begin{document}

\title{Isospin Symmetry of Fragmentation Functions}

\author{Kai-bao Chen}
\email{chenkaibao19@sdjzu.edu.cn}
\affiliation{School of Science, Shandong Jianzhu University, Jinan, Shandong 250101, China}

\author{Zuo-tang Liang}
\email{liang@sdu.edu.cn}
\affiliation{Institute of Frontier and Interdisciplinary Science, 
Key Laboratory of Particle Physics and Particle Irradiation (MOE), Shandong University, Qingdao, Shandong 266237, China}

\author{Yan-lei Pan}
\email{pyl@mail.sdu.edu.cn}
\affiliation{Institute of Frontier and Interdisciplinary Science, 
Key Laboratory of Particle Physics and Particle Irradiation (MOE), Shandong University, Qingdao, Shandong 266237, China}

\author{Yu-kun Song}
\email{sps\_songyk@ujn.edu.cn}
\affiliation{School of Physics and Technology, University of Jinan, Jinan, Shandong 250022, China}

\author{Shu-yi Wei}
\email{swei@ectstar.eu}
\affiliation{European Centre for Theoretical Studies in Nuclear Physics and Related Areas (ECT*)\\
and Fondazione Bruno Kessler, Strada delle Tabarelle 286, I-38123 Villazzano (TN), Italy}

\begin{abstract}
We make a systematic study of the isospin symmetry of fragmentation functions by taking decay contributions into account.  
We assume the isospin symmetry in strong interactions and show that in the unpolarized case the isospin symmetry is held for fragmentation functions of $\Lambda$  
and only tiny violations are allowed for other hadrons such as nucleon and pions due to the contributions from weak decays. 
We present a rough estimate of the magnitudes of such violations.
In the polarized case, we show that the isospin symmetry violation for $\Lambda$ production should be tiny 
and the recent Belle data on the transverse polarization of $\Lambda$ can be reproduced 
if the isospin symmetry is kept in the corresponding polarized fragmentation functions.
\end{abstract}

\maketitle

{\sc Introduction}: 
The fragmentation function (FF) is one of the most important quantities in describing hadron production in high energy reactions. 
The studies of FFs provide important information on the basic properties of quantum chromodynamics (QCD) and 
have been a standing topic in the field of High Energy Physics~\cite{Zyla:2020zbs,FFreviews}. 
Recent measurements on the transverse polarization of $\Lambda$ hyperon in $e^+e^-$ annihilations by Belle collaboration~\cite{Guan:2018ckx} 
open a new window in the studies and have attracted immediate attentions~\cite{Anselmino:2019cqd,DAlesio:2020wjq,Callos:2020qtu}. 
One of the most striking features of the Belle data~\cite{Guan:2018ckx} is the behaviors of the polarization of $\Lambda$ 
in the associated processes $e^+e^-\to \Lambda h^\pm X$ where $\Lambda$ and $h^\pm$ are in different hemispheres. 
The results show a distinct difference between the polarization in $e^+e^-\to \Lambda h^+ X$ and that in $e^+e^-\to \Lambda h^- X$, 
especially at small $z_\Lambda$, where $z_\Lambda$ is the longitudinal momentum fraction of the parent quark carried by $\Lambda$. 
Parameterizations have already been carried out by different groups~\cite{DAlesio:2020wjq,Callos:2020qtu} and the direct $\chi^2$-analysis 
seems to lead to a significant isospin symmetry violation in the corresponding polarized FF $D_{1Tq}^{\perp\Lambda}(z)$, i.e., $D_{1Tu}^{\perp\Lambda}(z) \neq D_{1Td}^{\perp\Lambda}(z)$. 

The isospin invariance is a fundamental property of QCD and strong interaction processes. 
The fragmentation process is dominated by the strong interaction. 
However, there can be contributions from electroweak interaction processes such as resonance decays. 
One therefore usually thinks that the isospin symmetry violation may be allowed to some extent in FFs 
and was often included in some of the parameterizations~\cite{FFreviews}. 
However, it has never been investigated up to what extent the violation can be accommodated. 
The parameterizations of the polarized FF $D_{1Tq}^{\perp\Lambda}(z)$ given in~\cite{DAlesio:2020wjq,Callos:2020qtu} 
show such a clear and substantial violation that could be a great challenge to QCD.  
It is therefore important and urgent to make a systematic study 
to see whether and, if yes, to what extent the isospin symmetry violation can be accommodated in FFs if it is held in strong interaction. 

Polarizations of hadrons produced in high energy reactions have been measured in different connections 
such as the transverse polarization of hyperons in $pp/pA$ collisions~\cite{Lesnik:1975my,Bunce:1976yb,Bensinger:1983vc,Gourlay:1986mf,Krisch:2007zza}, 
the spin transfer of the longitudinal polarization~\cite{Buskulic:1996vb,Ackerstaff:1997nh}, 
the spin alignment of vector mesons~\cite{Abreu:1997wd,Ackerstaff:1997kj,Ackerstaff:1997kd,Abbiendi:1999bz,Chukanov:2006xa} and 
the recent observation of the global polarization effect in $AA$ collisions~\cite{STAR:2017ckg,Liang:2004ph,Gao:2020lxh}. 
The results often bring us surprises and have inspired plenty of studies. 
For the hyperon polarization in fragmentation process, most of the studies concentrate on the polarized FF $G_{1Lq}^{\Lambda}(z)$, 
the longitudinal spin transfer~\cite{deFlorian:1997zj,Chen:2016iey}, and related phenomenological aspects~\cite{Gustafson:1992iq,Liang:1997rt,Dong:2004qs,Yang:2017cwi,Boros:1998kc,Liu:2000fi,Liu:2001yt,Liang:2002ub,Xu:2002hz,Ma:1998pd,Ma:2000uu,Chi:2013hka,Ellis:2002zv}. 
The Belle experiment has measured~\cite{Guan:2018ckx} the transverse polarization of $\Lambda$ in $e^+e^-$ annihilations. 
It is an induced polarization of hyperon produced in the fragmentation of an unpolarized quark 
and is described by the polarized FF $D_{1Tq}^{\perp \Lambda}(z)$. 
This is the first polarized FF of this class that has ever been parameterized~\cite{DAlesio:2020wjq,Callos:2020qtu}. 
The isospin symmetry has never been studied explicitly in this connection. 
It is therefore also important to clarify the situation for future studies.

In this paper, we present a systematic analysis of the isospin symmetry violation in FFs due to electroweak decays. 
We show that such violations can only be tiny as long as the isospin symmetry holds in strong interaction. 
We also demonstrate that the Belle data~\cite{Guan:2018ckx} can be fitted if the isospin symmetry for 
the polarized FF $D_{1Tq}^{\perp\Lambda}(z)$ is kept. 
There is no need to introduce a significant isospin symmetry violation in the FF.

{\sc The Unpolarized FF}:
We start with the fragmentation process $q\to h+X$ in the unpolarized case 
and write out decay contributions explicitly so that 
\begin{align}
&D_{1q}^{h}(z)=D_{1q}^{h,{\rm dir}}(z)+D_{1q}^{h,{\rm dec}}(z), \\
&D_{1q}^{h,{\rm dec}}(z)=\sum_{h_j} D_{1q}^{h,h_j}(z),\\
&D_{1q}^{h,h_j}(z)=Br(h,h_j) \int dz' K_{h,h_j}(z,z') D_{1q}^{h_j}(z'),  \label{eq:hjh}
\end{align}
where $D_{1q}^{h}(z)$ is the final FF; $D_{1q}^{h,{\rm dir}}(z)$ is the directly produced part;  
$D_{1q}^{h,{\rm dec}}(z)$ is the decay contribution part,   
$D_{1q}^{h,h_j}(z)$ is that from the channel $h_j\to hX$, 
$Br(h,h_j)$ is the branch ratio and $K_{h,h_j}(z,z')$ is the Kernel function representing the probability 
for a $h_j$ with momentum fraction $z'$ to decay into a $h$ with momentum fraction $z$. 

We limit ourselves to light flavors
and consider $J^P=0^-$ pseudo-scalar and $J^P=1^-$ vector mesons,  
and  $J^P=(1/2)^+$ octet and $J^P=(3/2)^+$ decuplet baryons. 
There is also small contribution from even higher excited resonance states~\cite{Hofmann:1988gy}. 
We do not consider them because: 
(1) the production rates are very small. 
(2) most of them decay via strong interaction so that the isospin symmetry is held in such decay processes. 
We just include them in the directly produced part effectively. 

The decay channels and the corresponding branch ratios can be found in the Review of Particle Properties~\cite{Zyla:2020zbs}. 
We analyze the situation for different hadrons one by one in the following. 

(i) {\it $\Xi$ production}: 
The situation for the baryon production is simpler than that for mesons since it receives decay contributions only from heavier baryons. 
We start with $\Xi$ where we have decay contributions from $\Xi^*$ and $\Omega^-$,  i.e., 
\begin{align}
&D_{1q}^{\Xi^{i}}(z)=D_{1q}^{\Xi^i,{\rm dir}}(z)+D_{1q}^{\Xi^i,\Xi^*}(z)+D_{1q}^{\Xi^i,\Omega^-}(z).
\end{align}
Here we use $i$ to specify one of the charge states and 
$D_{1q}^{\Xi^i,\Xi^*}(z)\equiv\sum_jD_{1q}^{\Xi^i,\Xi^{*j}}(z)$ to denote the sum of contributions of different charge states 
and similar in the following of this paper.
We take the isospin symmetry in strong interaction processes, and obtain,      
\begin{align}
&D_{1u}^{\Xi^0,{\rm dir}}(z)=D_{1d}^{\Xi^-,{\rm dir}}(z), \label{eq:Xi} \\
&D_{1u}^{\Xi^0,\Xi^*}(z)=D_{1d}^{\Xi^-,\Xi^*}(z),\\
&D_{1u}^{\Omega^-}(z)=D_{1d}^{\Omega^-}(z). \label{eq:Omega}
\end{align}
The weak decay of $\Omega^-$ leads indeed to a small isospin symmetry violation of $\Xi$ production given by
\begin{align}
\delta D_{1q}^{\Xi}(z)&\equiv D_{1u}^{\Xi^0}(z)-D_{1d}^{\Xi^-}(z)=\delta D_{1q}^{\Xi,\Omega^-}(z) \nonumber\\
&=\delta Br(\Xi,\Omega^-)\int dz'K_{\Xi,\Omega^-}(z,z')D_{1u}^{\Omega^-}(z'), \label{eq:dXi}
\end{align}
because there is a difference between the branch ratios of $\Omega^-\to \Xi^0\pi^-$ and $\Omega^-\to\Xi^-\pi^0$,
$\delta Br(\Xi,\Omega^-)\equiv Br(\Xi^0,\Omega^-)-Br(\Xi^-,\Omega^-)=0.150$~\cite{Zyla:2020zbs}. 
However, the effect is tiny because the total contribution from $\Omega^-$ decay to $\Xi$ in the final state is only a very small fraction.

(ii) {\it $\Lambda$ production}: 
The $\Lambda$ production receives contributions from $\Omega^-$, $\Sigma^*$, $\Xi$ and $\Sigma^0$ decays, i.e.,  
\begin{align}
D_{1q}^{\Lambda}(z)=&D_{1q}^{\Lambda, {\rm dir}}(z)+D_{1q}^{\Lambda,\Omega^-}(z)+D_{1q}^{\Lambda,\Sigma^*}(z)\nonumber\\
&+D_{1q}^{\Lambda,\Xi}(z)+D_{1q}^{\Lambda,\Sigma^0}(z).
\end{align}

With the isospin symmetry for strong interaction contributions,  we have, similar to Eqs.~(\ref{eq:Xi})-(\ref{eq:Omega}), that  
\begin{align}
&D_{1u}^{\Lambda,{\rm dir}}(z)=D_{1d}^{\Lambda,{\rm dir}}(z),\\
&D_{1u}^{\Lambda,\Sigma^*}(z)=D_{1d}^{\Lambda,\Sigma^*}(z),\\
&D_{1u}^{\Sigma^0}(z)=D_{1d}^{\Sigma^0}(z). \label{eq:Sigma0}
\end{align}

From Eqs.~(\ref{eq:Omega}) and (\ref{eq:Sigma0}), we obtain immediately that there is no contribution to isospin symmetry violation in $\Lambda$ production 
from $\Omega^-\to\Lambda X$ or $\Sigma^0\to\Lambda\gamma$ because the decay process involved is the same for $u\to\Lambda +X$ and $d\to\Lambda +X$.  

For the contribution from the weak decay of $\Xi$, 
although there is a small isospin symmetry violation for $\Xi$ production, i.e., a difference between $D_{1u}^{\Xi^0}(z)$ and $D_{1d}^{\Xi^-}(z)$ as given by Eq.~(\ref{eq:dXi})
due to the weak decay contribution from $\Omega^-$, it does not lead to isospin symmetry violation in the $\Lambda$ production.  
This is because both $\Xi^0$ and $\Xi^-$ contribute equally to $\Lambda$ production via the weak decay $\Xi\to\Lambda\pi$. 
Such a difference disappears in the $\Lambda$ production where the contributions 
from $\Xi^0\to\Lambda\pi^0$ and $\Xi^-\to\Lambda\pi^-$ add together, i.e., we still obtain
\begin{align}
&D_{1u}^{\Lambda,\Xi}(z)=D_{1d}^{\Lambda,\Xi}(z).
\end{align}

We conclude that the electroweak decays do not lead to isospin symmetry violation in $\Lambda$ production 
and the isospin symmetry is valid in unpolarized FFs of $\Lambda$, i.e., 
\begin{align}
&D_{1u}^{\Lambda}(z)=D_{1d}^{\Lambda}(z).
\end{align}

We note in particular that this contradicts the parameterizations in e.g. \cite{Albino:2008fy} 
where significant isospin symmetry violations are involved in FFs of $\Lambda$.

(iii) {\it Nucleon production}: 
Nucleon receives the most decay contributions among all the different baryons. 
There are weak decay contributions from $\Lambda$ and $\Sigma^\pm$.  
Both of them lead to isospin symmetry violation. 
We denote it by  $\delta D_{1q}^N\equiv D_{1u}^{p}-D_{1d}^{n}$, and obtain 
\begin{align}
&\delta D_{1q}^{N}(z)=\delta D_{1q}^{N,\Lambda}(z)+\delta D_{1q}^{N,\Sigma}(z), 
\end{align}
The corresponding decay branch ratio differences are~\cite{Zyla:2020zbs} $\delta Br(N,\Lambda)=0.281$ and $\delta Br(N,\Sigma)=-0.484$.

(iv) {\it $\pi$ production:} 
Among all different hadrons, the production of pions receives most decay contributions and the situation is most complicated. 

Besides strong decays, we have contributions from electroweak decays of the isospin singlet mesons $\omega$, $\phi$, $\eta$ and $\eta'$
and these isospin singlet meson decays do not lead to isospin symmetry violation for pion production. 

There is contribution to isospin symmetry violation from decays of kaons. 
There are also decay contributions from baryons and anti-baryons. 
Among them,  we have contributions to isospin symmetry violations from weak decays of $\Omega^-,\Lambda,\Xi^-$, so that 
\begin{align}
&\delta D_{1q}^{\pi}(z)=\delta D_{1q}^{\pi,K}(z)+\delta D_{1q}^{\pi,B}(z)+\delta D_{1q}^{\pi,\bar B}(z).
\end{align}

(v) {\it A rough estimate}:
We see that weak decay contributions lead indeed to isospin symmetry violations in FFs for productions of hadrons such as $\Xi$, nucleon and pion. 
A precise calculation of such violations is, however, quite involved and depends on FFs of different hadrons. 
We make a rough estimate of magnitudes in the following.  

We consider a system created in fragmentations of quarks and anti-quarks. 
We assume equal numbers of $u$ and $d$ but a strangeness suppression factor $\lambda$ for $s$; and equal numbers of quarks and anti-quarks of the same flavors. 
We assume isospin symmetry for directly produced hadrons but the same suppression $\lambda$ for hadrons with a strange quark. 
In this way, each hadron acquires a relative production weight and we calculate the ratios of average yields in the final state as a guide of the magnitudes of 
the isospin symmetry violations. 

For $\Xi$ and nucleon, we calculate the difference between the average yields of 
$\Xi^0$ and $\Xi^-$ and that of protons and neutrons. 
For $\Xi$, this is given by
\begin{align}
&{\delta \langle N_\Xi\rangle}/{\langle N_\Xi\rangle} =({\langle N_{\Xi^0}\rangle-\langle N_{\Xi^-}\rangle})/(\langle N_{\Xi^0}\rangle+\langle N_{\Xi^-}\rangle)
\nonumber\\
&~~~=({0.150\lambda^3\gamma})/[{2\lambda^2+\gamma(2\lambda^2+0.322\lambda^3)}], 
\end{align}
where $\gamma$ is the relative production weight of  $J^P=(1/2)^+$ to $J^P=(3/2)^+$ baryon.  
We take $\lambda=0.3$ and $\gamma=0.3$ as obtained~\cite{Hofmann:1988gy} by fitting the data available, 
and obtain 
${\delta \langle N_\Xi\rangle}/{\langle N_\Xi\rangle}\approx 0.005$. 
Similarly, for nucleon, we obtain ${\delta \langle N_N\rangle}/{\langle N_N\rangle}\approx 0.026$.
Both are less than a few percent.

For $\pi$, we estimate the largest contribution from the kaon decay 
by calculating the contribution of one pion from the kaon decay and the relative contribution from kaon decay to the production of pion. 
The former is given by the ratio $\delta Br(\pi,K)/Br(\pi,K)=-0.26$~\cite{Zyla:2020zbs}  
where $\delta Br(\pi, K)=Br(\pi^+,K^+)-Br(\pi^-,K^0)$ and $Br(\pi, K)=Br(\pi^+,K^+)+Br(\pi^-,K^0)$. 
For the latter, we obtain ${\langle N_{\pi^+}^{\pi^+,K}\rangle}/{\langle N_{\pi^+}\rangle}\sim 0.16$ 
by taking the relative weight of vector to pseudo-scalar meson  $\beta\sim 1$ and neglecting decay contributions from baryons and anti-baryons. 
Together this gives a violation of less than 5\%.

{\sc The Polarized FFs of $\Lambda$}:
We have shown that weak decays do not lead to isospin symmetry violation of unpolarized FFs of $\Lambda$. 
For polarized FFs, we need to consider the spin transfer in the corresponding decay process.  
It is obvious that the decays of $\Omega^-\to\Lambda\pi^-$ and $\Sigma^0\to\Lambda\gamma$ 
do not lead to isospin symmetry violation in FFs of $\Lambda$ production 
because they are the same for $u\to\Lambda +X$ and $d\to\Lambda +X$.  

The only candidate for the violation is the contribution from the weak decay $\Xi\to\Lambda\pi$. 
The spin transfer in this decay process $\Xi\to\Lambda\pi$ depends on the direction of momentum of $\Lambda$ in the rest frame of $\Xi$ 
and is determined by the decay parameters. The average over all directions is $(1+2\gamma)/3$. 
The decay parameters can be found in the Review of Particle Properties~\cite{Zyla:2020zbs}. 
We see indeed some differences between the decay parameters for $\Xi^0\to\Lambda\pi^0$ and those for $\Xi^-\to\Lambda\pi^-$. 
Such differences lead to the isospin symmetry violation in polarized FFs of $\Lambda$ production. 
But the difference is not large, e.g., $\gamma=0.85$ for $\Xi^0\to\Lambda\pi^0$ and is $0.89$ for $\Xi^-\to\Lambda\pi^-$, 
leading to a difference of $2\delta\gamma/3\sim 0.027$ in the spin transfer. 
Positivity bounds demand that the violation should be, in any case, less than this factor times $D_{1u}^{\Lambda,\Xi}(z)$.  
Having in mind that $D_{1u}^{\Lambda,\Xi}(z)$ is only a small fraction of $D_{1u}^{\Lambda}(z)$, 
we conclude that the isospin symmetry violation in the polarized FFs of $\Lambda$ can only be a tiny effect. 

\begin{figure}[tb]
\includegraphics[width=0.45\textwidth]{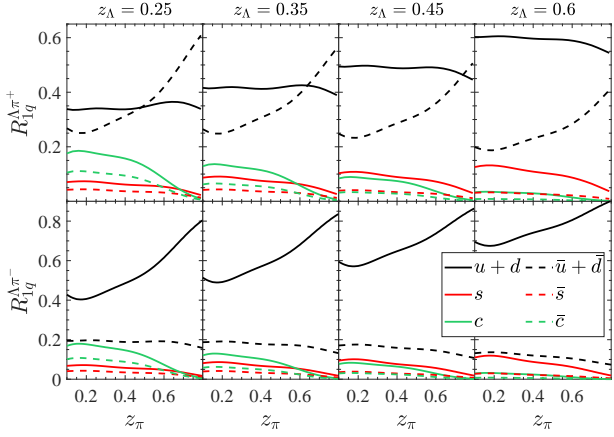}
\caption{The relative weight of different flavors in $e^+e^-\to\Lambda\pi X$ 
calculated with DSV~\cite{deFlorian:1997zj} and DHESS~\cite{deFlorian:2014xna} FFs.}
\label{fig:Rqpi}
\end{figure}

{\sc The fit to the Belle data}:
Now we just assume the isospin symmetry for the polarized FF $D_{1Tq}^{\perp\Lambda}$ and study 
whether we can describe the  Belle data~\cite{Guan:2018ckx}. 
We use the Trento convention~\cite{Bacchetta:2004jz} for the transverse momentum dependent FF $D_{1Tq}^{\perp\Lambda} (z, p_T)$ 
and take a factorized form $D_{1Tq}^{\perp\Lambda} (z,p_T) = D_{1Tq}^{\perp\Lambda} (z) \frac{1}{\pi \Delta^2}\exp(-p_T^2/\Delta^2)$, 
where the transverse momentum dependent part is taken as a Gaussian with a flavor independent width $\Delta$. 
The Belle experiment measured the $p_T$-integrated polarizations in the inclusive process $e^+e^-\to \Lambda X$, where the transverse direction is defined with respect to the thrust axis, 
and the associated process $e^+e^-\to \Lambda h X$, where $h$ and $\Lambda$ are in different hemispheres. 
They are given by 
\begin{align}
P_{\Lambda} & (z_\Lambda) = \frac{\sqrt{\pi}}{2} \frac{\Delta}{z_\Lambda M_\Lambda} \frac{\sum_q e_q^2 D_{1Tq}^{\perp\Lambda} (z_\Lambda)}{\sum_q e_q^2 D_{1q}^\Lambda (z_\Lambda)},\\
P_{\Lambda}&(z_\Lambda,z_h)
= \frac{\sqrt{\pi}}{2} \frac{z_h\Delta}{z_\Lambda M_\Lambda\sqrt{z_h^2 + z_\Lambda^2\Delta_h^2/\Delta^2}} \nonumber \\
& \times \sum_q \Big[R_{1q}^{\Lambda h} (z_\Lambda,z_h) \frac{D_{1Tq}^{\perp\Lambda}(z_\Lambda)} {D_{1q}^{\Lambda}(z_\Lambda)} +(q\leftrightarrow\bar q)\Big], 
\label{eq:PLambda}
\end{align}
respectively. 
Here $R_{1q}^{\Lambda h} (z_\Lambda,z_h)$ is the relative weight of the contribution from $q\to\Lambda X$ and $\bar q\to hX$.
At the leading order in pQCD, it is given by
\begin{align}
R_{1q}^{\Lambda h} (z_\Lambda,z_h) = \frac{ e_q^2 D_{1q}^{\Lambda}(z_\Lambda) D_{1\bar q}^{h}(z_h)}{\sum_f  e_f^2 D_{1f}^{\Lambda}(z_\Lambda) D_{1\bar f}^{h}(z_h)+(f\leftrightarrow\bar f)}.
\end{align}
We see that $R_{1q}^{\Lambda h}(z_\Lambda,z_h)$ is determined by unpolarized FFs. 
We calculate it by using different parameterizations~\cite{deFlorian:1997zj,deFlorian:2007aj,Albino:2008fy,deFlorian:2014xna,deFlorian:2017lwf} and {\sc Pythia}~\cite{Sjostrand:2000wi}. 
The results obtained are slightly different from each other but the qualitative features are the same.  
We show the results obtained using DSV~\cite{deFlorian:1997zj} for $\Lambda$ and DHESS~\cite{deFlorian:2014xna} for $\pi$ in Fig.~\ref{fig:Rqpi}.

From Fig.~\ref{fig:Rqpi}, we see clearly that there is a significant difference between $R_{1q}^{\Lambda\pi} (z_\Lambda,z_\pi)$ in $e^+e^-\to\Lambda\pi^+ X$ 
and that in $e^+e^-\to\Lambda\pi^- X$. 
Such a difference is more obvious at smaller $z_\Lambda$ but decreases with increasing $z_\Lambda$. 
It is also obvious that such a difference might be the source for the different behaviors of $\Lambda$ polarizations in $e^+e^-\to\Lambda\pi^\pm X$
observed in the Belle experiment~\cite{Guan:2018ckx}. 

With the isospin symmetry for $D_{1Tq}^{\perp\Lambda}$, we have $D_{1Tu}^{\perp\Lambda}= D_{1Td}^{\perp\Lambda}$ 
and $D_{1T\bar u}^{\perp\Lambda}= D_{1T\bar d}^{\perp\Lambda}$. 
However, since $s$ contributes quite differently to the spin of $\Lambda$, we expect $D_{1Ts}^{\perp\Lambda}$ to be quite different from $D_{1Tu}^{\perp\Lambda}=D_{1Td}^{\perp\Lambda}$. 
At the Belle energies~\cite{Guan:2018ckx}, a significant contribution to $\Lambda$ comes from $c$ and $\bar c$ fragmentation (c.f. Fig.~\ref{fig:Rqpi}). 
The heavy quarks fragment quite differently from light quarks and the leading hadron contribution in $c\to\Lambda X$ comes mainly from the decay $\Lambda_c\to \Lambda X$. 
We therefore expect that $D_{1Tc}^{\perp\Lambda}$ and $D_{1T\bar c}^{\perp\Lambda}$ can also be quite different from those of light quarks.

\begin{figure}[htb]
\includegraphics[width=0.45\textwidth]{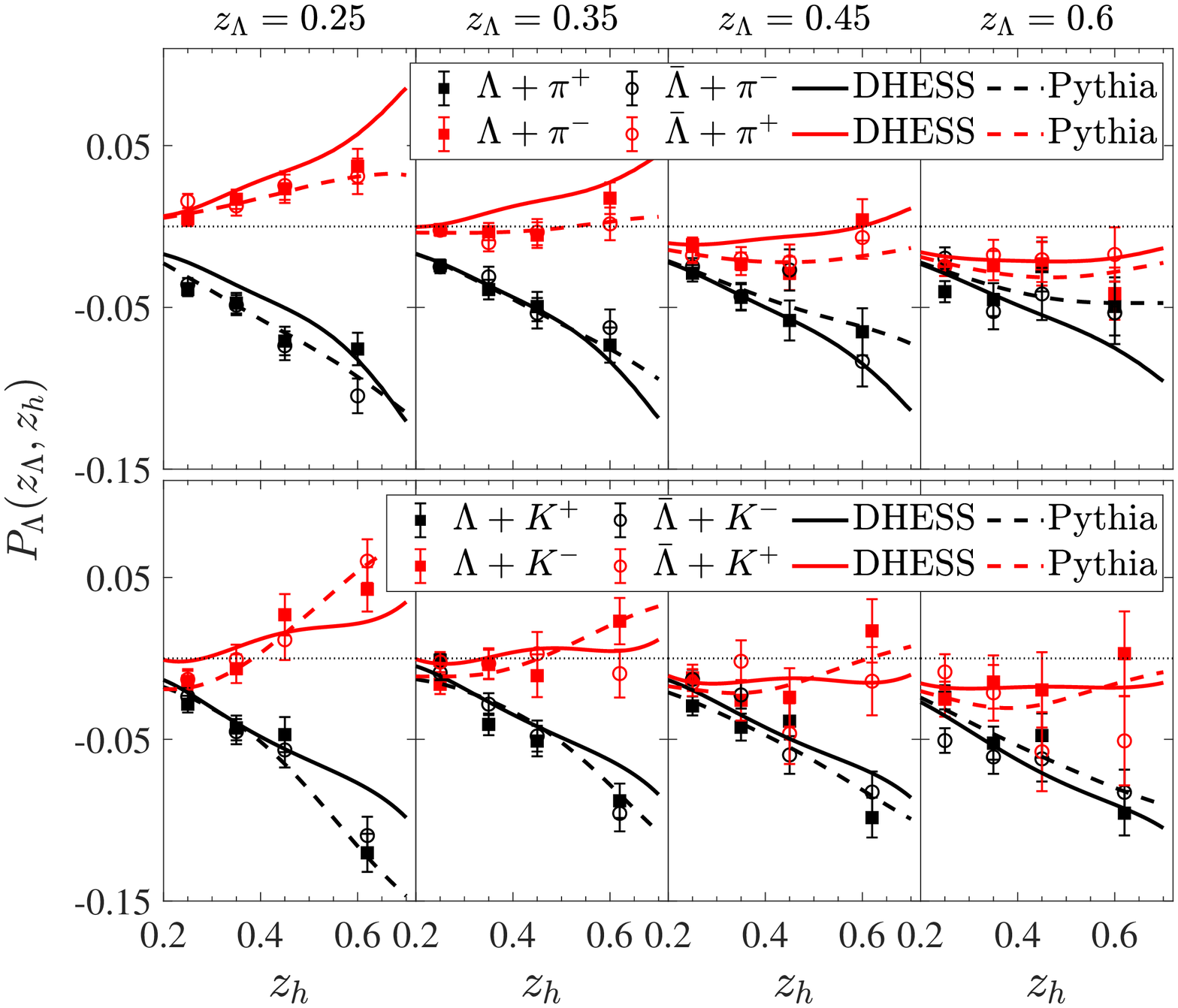}
\caption{Transverse polarization of $\Lambda$ in $e^+e^-\to\Lambda h^\pm X$ compared with the Belle data~\cite{Guan:2018ckx}. 
}
\label{fig:polarization}
\end{figure}

\begin{figure}[htb]
\includegraphics[width=0.35\textwidth]{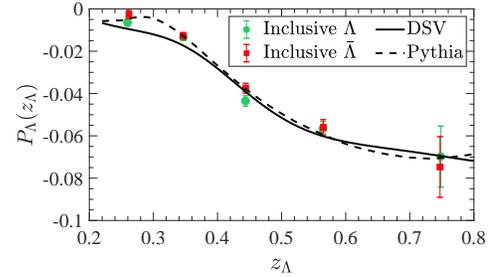}
\caption{Transverse polarization of $\Lambda$ in $e^+e^- \to \Lambda X$ compared with the Belle data~\cite{Guan:2018ckx}. 
}
\label{fig:polarization-inclusive}
\end{figure}

We parametrize $D_{1Tq}^{\perp\Lambda}(z)$ with the same form as that used in~\cite{DAlesio:2020wjq,Callos:2020qtu}, i.e., 
\begin{align}
D_{1Tq}^{\perp\Lambda}(z)=&N_{Tq} \frac{(\alpha_{Tq}+\beta_{Tq}-1)^{\alpha_{Tq}+\beta_{Tq}-1} } {(\alpha_{Tq}-1)^{\alpha_{Tq}-1}\beta_{Tq}^{\beta_{Tq}} } \nonumber\\
&\times z^{\alpha_{Tq}} (1-z)^{\beta_{Tq}} D_{1q}^{\Lambda}(z). 
\end{align}
The isospin symmetry ensures that $N_{Tq}$, $\alpha_{Tq}$ and $\beta_{Tq}$ take the same values for $q = u$ and $d$ and for $\bar q=\bar u$ and $\bar d$. 
For other quarks and antiquarks, they are just free parameters to be extracted with a $\chi^2$-analysis to the Belle data~\cite{Guan:2018ckx}. 
The $\chi^2$ finds its minimum with the set of values for the free parameters listed in Table~\ref{tab:parameters}. 
We show our results of the transverse polarization of $\Lambda$ compared with the Belle data~\cite{Guan:2018ckx} 
in Figs.~\ref{fig:polarization} and~\ref{fig:polarization-inclusive}. 

\begin{table} [b]\centering
\caption{Values of parameters extracted with a $\chi^2$-analysis to the Belle data~\cite{Guan:2018ckx}. 
The first and second rows present the results with inputs of unpolarized FFs from 
DSV+DHESS unpolarized FFs~\cite{deFlorian:1997zj,deFlorian:2014xna,deFlorian:2017lwf}
and {\sc Pythia}~\cite{Sjostrand:2000wi} respectively. 
The corresponding values of $\Delta^2/\Delta_h^2$ are $0.951$ and $0.939$.} \label{tab:parameters}
\begin{tabular}{c|cccccc} \hline\hline
 parameter  
& $u, d$ 
& $s$ 
& $c$
& $\bar u,\bar d$
& $\bar s$
& $\bar c$ \\ \hline
\multirow{2}{*}{$\frac{\Delta}{M_\Lambda} N_{Tq}$}
& ~$0.391$~
& $-0.391$
& $0.0278$
& $-0.456$
& $-0.430$
& ~$0.401$~
\\
& $0.245$
& $-0.148$
& $0.108$
& $-0.231$
& $0.523$
& $-0.324$
\\ \hline
\multirow{2}{*}{$\alpha_q$}
& $1.38$
& $6.91$
& $1.43$ 
& $1.00$
& $2.64$
& $11.6$
\\
& $2.41$
& $1.54$
& $5.14$
& $1.86$
& $1.74$
& $1.02$
\\ \hline
\multirow{2}{*}{$\beta_q$}
& $3.98$
& $0.646$
& $14.3$
& $0.0319$
& $2.77$
& $14.9$
\\
& $7.69$
& $0.551$
& $15.0$
& $2.35$
& $14.9$
& $2.41$
\\ \hline\hline
\end{tabular}
\end{table}

From Figs.~\ref{fig:polarization} and~\ref{fig:polarization-inclusive}, we see that the Belle data~\cite{Guan:2018ckx} 
are well described under the constraint of isospin symmetry for the corresponding polarized FFs. 
Furthermore, we also see that there are significant differences between the parameters extracted 
with DSV+DHESS unpolarized FFs~\cite{deFlorian:1997zj,deFlorian:2014xna,deFlorian:2017lwf} 
and those with {\sc Pythia}~\cite{Sjostrand:2000wi}. 
This shows the uncertainties in unpolarized FFs have also quite large influences on the results of $\Lambda$ polarizations. 
The data available~\cite{Guan:2018ckx} are still far from enough to determine the corresponding polarized FFs accurately. 
There is no need to introduce significant isospin symmetry violation in FFs at this stage. 

{\sc Summary and outlook}:
We made a systematic study of the isospin symmetry of FFs by taking resonance decays into account. 
The results show that there is no reason to introduce significant isospin symmetry violation in FFs 
if we assume that isospin symmetry is valid in strong interactions. 
We see in particular that the isospin symmetry is expected to be valid in unpolarized FFs for $\Lambda$ hyperon production    
and in the polarized case there can only be a tiny violation due to the contribution from $\Xi\to\Lambda\pi$. 
We demonstrated that, in contrast with the recent parameterizations~\cite{DAlesio:2020wjq,Callos:2020qtu}, 
the Belle data~\cite{Guan:2018ckx} on the transverse polarization of $\Lambda$ 
can also be well described without isospin symmetry violation in the parameterization of corresponding FFs.

The results demand that isospin symmetry violation should be dealt with carefully in parameterizing FFs. 
In particular for $\Lambda$ production, we need to take it as granted in unpolarized FFs. 
In the polarized case, we may include a tiny violation due to the contribution from $\Xi$ decays 
with an upper limit of a few percent of the contribution from the $\Xi$ decay to the unpolarized FF.  

{\sc Acknowledgements}:
This work was supported in part by the National Natural Science Foundation of China
(approval Nos. 11890713, 1205122, 11890710, 11947055, 11505080) and Shandong Province Natural Science Foundation Grant Nos. ZR2018JL006 and ZR2020QA082.

\end{document}